# Phase-Coherent Charge Transport through a Porphyrin Nanoribbon

Zhixin Chen[1*†], Jie-Ren Deng[2†], Songjun Hou[3], Xinya Bian[1], Jacob L. Swett[1], Qingqing Wu[3], Jonathan Baugh[4], L. Bogani[1], G. Andrew D. Briggs[1], Jan A. Mol[5], Colin J. Lambert[3*], Harry L. Anderson[2*], and James O. Thomas[1*]

[1]Department of Materials, University of Oxford, Parks Road, Oxford, OX1 3PH, UK

[2]Department of Chemistry, University of Oxford, Chemistry Research Laboratory, Oxford, OX1 3TA, UK

[3]Department of Physics, Lancaster University, Lancaster, LA1 4YB, UK

[4]Institute for Quantum Computing, University of Waterloo, Waterloo, ON N2L 3G1, Canada

[5]School of Physical and Chemical Sciences, Queen Mary University, London, E1 4NS, UK

*zhixin.chen@materials.ox.ac.uk,     *c.lambert@lancaster.ac.uk,     *harry.anderson@chem.ox.ac.uk, *james.thomas@materials.ox.ac.uk     † These authors made equal contributions.

**Quantum interference in nano-electronic devices could lead to reduced-energy computing and efficient thermoelectric energy harvesting. When devices are shrunk down to the molecular level it is still unclear to what extent electron transmission is phase coherent, as molecules usually act as scattering centres, without the possibility of showing particle-wave duality. Here we show electron transmission remains phase coherent in molecular porphyrin nanoribbons, synthesized with perfectly defined geometry, connected to graphene electrodes. The device acts as a graphene Fabry-Pérot interferometer, allowing direct probing of the transport mechanisms throughout several regimes, including the Kondo one. Electrostatic gating allows measurement of the molecular conductance in multiple molecular oxidation states, demonstrating a thousand-fold increase of the current by interference, and unravelling molecular and graphene transport pathways. These results demonstrate a platform for the use of interferometric effects in single-molecule junctions, opening up new avenues for studying quantum coherence in molecular electronic and spintronic devices.**

## Introduction

The ability to harness and exploit coherence at the nanoscale is crucial for emerging quantum technologies being developed in research areas across engineering, chemistry, and condensed-matter physics.[1] Single-molecule devices are an excellent platform to study quantum-coherent phenomena because molecular structures are atomically defined,[2] and recently, bottom-up synthesised graphene[3] and molecular nanoribbons[4] have attracted considerable attention in quantum information processing due to their low dimensionality and associated topological states. Most studies of single molecules in junctions have focussed on observing quantum phenomena, such as interference, in two-terminal devices by comparing transport properties of homologous series of molecules[5,6] and averaging data such that details of different transport mechanisms could be lost, rather than manipulating and studying QI within the same molecule.[7,8] The ability to measure and tune transport properties of the same single-molecule device, whilst changing gate potential, magnetic field and temperature, is necessary to understand how different transport mechanisms that arise from molecular, electrode, and molecule-electrode hybrid states all come together to contribute to device conductance. Studies of this nature can answer questions such as: how can electron transmission be shown to be phase coherent? What molecule-electrode coupling regime[9] is required for this to be the case? Moreover, the understanding gained from these studies feeds into one of the ultimate goals of single-molecule electronics, which is how to integrate molecules and nanoribbons,[10] one-by-one, into solid-state devices with some functionality that exploits the quantum properties of an individual molecule.

Molecular devices that use graphene as an electrode material are an ideal platform for investigating phase-coherent phenomena in charge transport.[11-13] The spatial confinement and long coherence lengths of electrons in graphene mean that devices display a range of quantum-coherent features, such as electronic Fabry-Pérot interferometry.[14,15] Furthermore, there are established routes to interface molecules with graphene through π-stacking interactions[16] to generate electrostatically gated single-molecule graphene junctions, enabling temperature and magnetic-field dependent measurements with a high operating frequency.[9,17] This is facilitated by the weaker screening by graphene of the gate electric field compared with bulky 3D metallic electrodes used in traditional molecular junctions[18,19] and by the electronic band structure of graphene, which makes it possible to differentiate between the contributions to transport of graphene and molecular states.[20]

In this work, we study charge transport through porphyrin nanoribbon-graphene devices at cryogenic temperatures (Fig. 1a) and demonstrate that conductance measurements as a function of (source-drain) bias voltage ($V_{sd}$), gate voltage ($V_g$), magnetic field, and temperature exhibit a wide range of QI phenomena such as Fabry-Pérot and Kondo resonances. Our results demonstrate that an 8-nm porphyrin octamer nanoribbon sustains phase-coherent electron transmission that can be tuned electrostatically, and is highly oxidation-state dependent, and demonstrates a comprehensive picture of the different quantum-coherent phenomena that arise in molecule-graphene junctions.

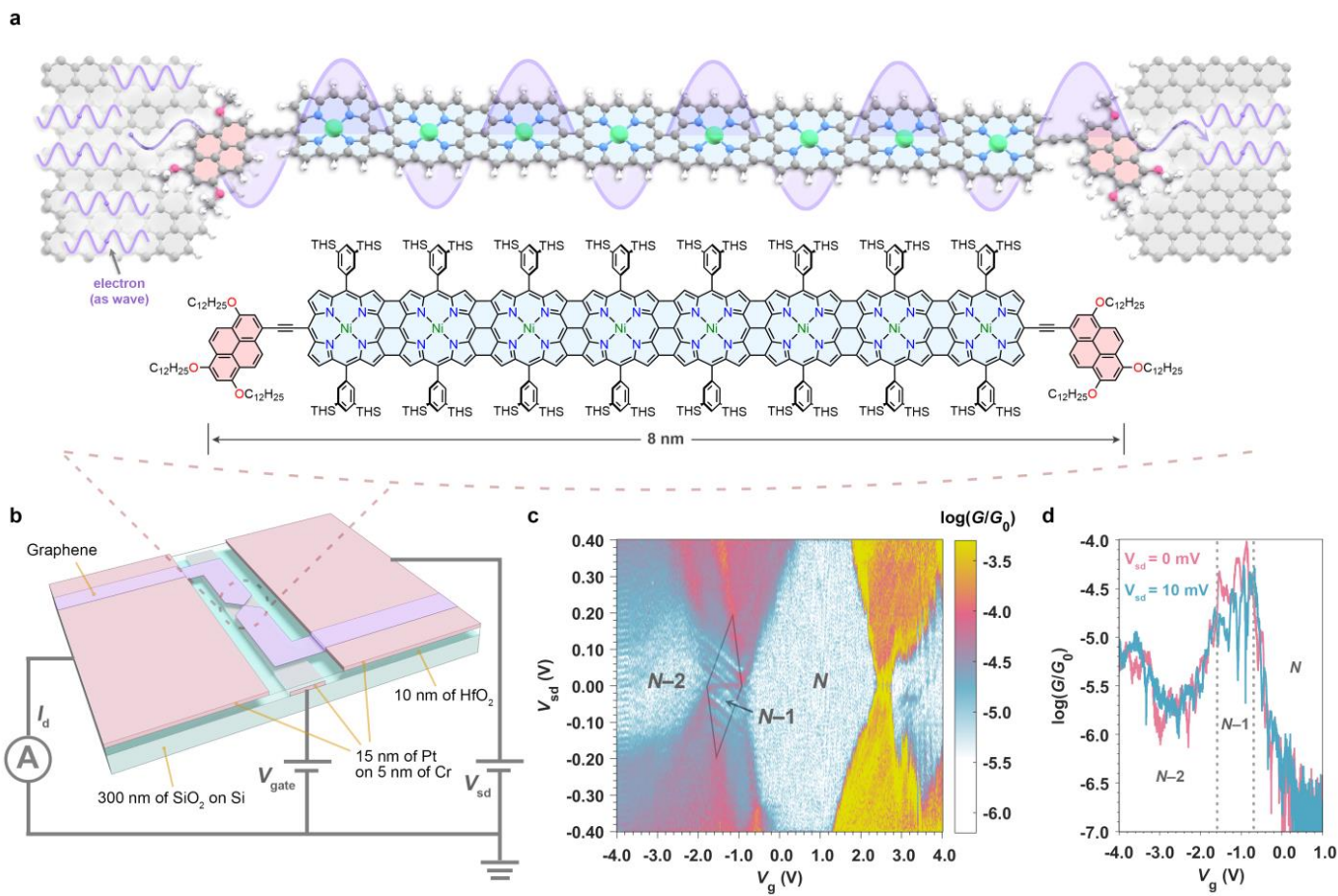

**Fig. 1 | Phase-coherent electron transport through a porphyrin nanoribbon-graphene device. a** Schematic representation of phase-coherent electron transport through a graphene-**Ni-FP8** porphyrin nanoribbon-graphene junction (top) and full chemical structure of **Ni-FP8** (bottom, THS = trihexylsilyl). In the top panel, solubilising groups on porphyrins (3,5-*bis*(trihexylsilyl)phenyl) and pyrenyls (dodecyloxy) have been omitted for clarity. **b** Device architecture. The pink rectangular area in the middle represents the local platinum gate electrode under a 10 nm layer of $HfO_2$ (transparent light blue); the rectangular areas (pink) at both ends represent the source and drain platinum electrodes, which are in contact with the bowtie-shape graphene (purple). **c** Differential conductance ($G = dI_{sd}/dV_{sd}$) map measured as a function of bias voltage ($V_{sd}$) and gate voltage ($V_g$) and **d** differential conductance as a function of $V_g$ at $V_{sd} = 0$ mV (pink curve) and $V_{sd} = 10$ mV (blue curve) for **Ni-FP8** device at 4.2 K. The conductance is plotted in logarithmic scale as the ratio to conductance quantum $G_0 = 2e^2/h$ where $e$ is the elementary charge and $h$ is Planck's constant. The $N–1$ charge state is highlighted by a grey diamond.

## Results

### Fabrication of single porphyrin-nanoribbon//graphene junctions

We focus on charge transport through an edge-fused nickel(II) porphyrin octamer (**Ni-FP8**) device (Fig. 1a). **Ni-FP8** has a length of 8 nm, and we term it a porphyrin nanoribbon, because it has a similar aspect ratio and electronic structure to a graphene nanoribbon.[21,22] The synthesis of the **Ni-FP8** nanoribbon is described in Methods Section, Fig. 1 in Extended Data and Supplementary Information Section 2. Most previous transport studies with porphyrin oligomers have used zinc(II) complexes,[23,24] but in this case nickel was used to facilitate the synthesis and to reduce the energy of the HOMO, thus enhancing the chemical stability of the ribbon. The fused porphyrin octamer core is functionalised with pyrene anchor groups to bind to the graphene source and drain electrodes via π-stacking.

A schematic of the device used in this study is shown in Fig. 1b. The architecture comprises nanometre-spaced graphene source and drain electrodes and a platinum electrostatic gate electrode. The gate, covered with a 10 nm $HfO_2$ dielectric layer, tunes the chemical potentials of the molecular states and the graphene. The device design allows multiple molecular charge states to be measured due to a strong electrostatic coupling between the gate voltage and the molecular states (the gate coupling, $\alpha_g$, is typically larger than 0.1 eV/V), and the high dielectric breakdown voltage of $V_g \approx \pm 5$ V. We have optimised a previous fabrication procedure to use a Z-shaped pattern for the graphene to reduce tension in the constriction, and use a positive, rather than negative, photoresist to reduce contamination (details of this are given in the Methods section). A combination of these two changes may contribute to the stronger electronic coupling we observe in this work compared to previous studies.[16] Although we discuss transport through a **Ni-FP8** device in detail, we also present data from another **Ni-FP8** device in Extended Data Fig. 2 and two from shorter porphyrin oligomers: one with a zinc porphyrin monomer (**Zn-P1**), one with an edge-fused porphyrin trimer (**Zn-FP3**).

**Charge transport measurement**

The full differential conductance map ($G_{sd} = dI_{sd}/dV_{sd}$) of the porphyrin nanoribbon device, measured at 4.2 K, displays several Coulomb diamonds and associated resonant tunnelling regions (Fig. 1c). We assign the diamond at $V_g = 0$ as the $N$ state ($N$ being the number of electrons on the molecule when it is neutral), as this diamond has a large addition energy, at $E_{add} = 0.7$ eV, that broadly matches the optical HOMO-LUMO gap (energy of the longest wavelength absorption band; 1750 nm, 0.71 eV). There is a zero-bias conductance peak in the neighbouring diamond between $V_g \approx -0.7$ V and $V_g \approx -1.4$ V, this peak is consistent with the experimental signature of a Kondo resonance (discussed in more detail below) that results from screening of an unpaired spin on the nanoribbon by electrons within the graphene electrodes.[25,26] The observation of a Kondo resonance is consistent with assigning this smaller diamond to the odd $N-1$ state (i.e. molecule is the radical cation, **Ni-FP8$^+$**, in this Coulomb diamond), confirming the larger diamond at $V_g = 0$ is the even $N$ state, as the number of electrons on the molecule differs by one between adjacent Coulomb diamonds.[27] The two sequential transport regions with broad edges which flank the $N-1$ diamond are then the $N-1/N$ charge transition ($V_g \sim -0.7$ V) and $N-2/N-1$ transition ($V_g \sim -1.4$ V). Finally, the highly conductive region at positive $V_g$ is the $N/N+1$ charge transition. We calculate the coupling of the molecular levels to the gate and source potentials from the slopes of the Coulomb diamonds, as $\alpha_{g,mol} = 0.22$ eV/V, and $\alpha_{s,mol} = 0.65$ eV/V respectively, giving the fraction of each applied potential that the molecular levels shift by.[19]

Considering the low measurement temperature (4.2 K, $k_BT = 0.4$ meV), the poor definition in the boundaries of the Coulomb diamonds (FWHM of Coulomb peaks ~ 14 meV, see Supplementary Figure S4-1) is attributed to lifetime broadening that results from intermediate molecule-electrode coupling, a regime consistent with appearance of the Kondo resonance.[28,29] The broad diamond edges suggest there are large regions where the molecular charge state is not well defined, in contrast with a device in the weakly coupled regime (commonly observed for molecules π-stacked to graphene electrodes[16,19,30]), where conductance occurs only within the sequential transport regions (when the chemical potentials of molecular transitions lie within the bias window, neglecting coherent resonant tunnelling) separated by Coulomb diamonds. In Fig. 1c and 1d we observe off-resonance transport features showing that, even at $V_{sd} = 10$ mV (away from the Kondo peak), the conductance is not completely suppressed. Instead, conductance remains above the noise except within the $N$ diamond, indicating that there are significant contributions from off-resonant phase-coherent transport around the $N-1/N$ and $N-2/N-1$ transitions of the intermediately coupled nanoribbon device.[28] As the hybridisation between molecule and electrode increases with electronic coupling, transport through an intermediately coupled molecular junction can only be fully understood by considering the entire graphene-nanoribbon-graphene system. We discuss the mixture of effects that arise from this holistic approach by initially focussing on phase-coherent transport within the graphene channel and then on molecular transport.

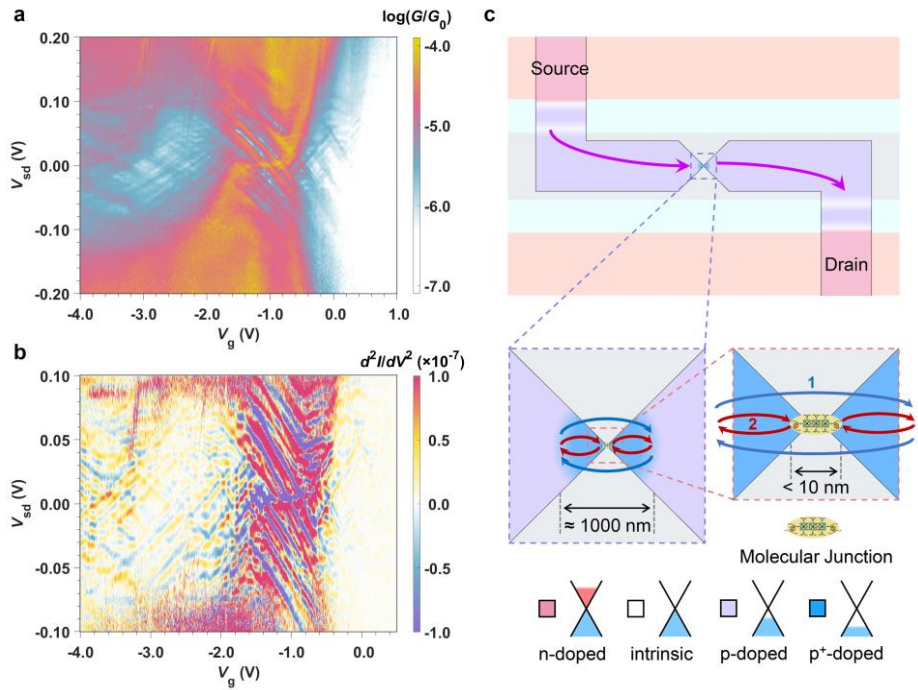

**Fig. 2 | Fabry-Pérot interference. a** Detailed conductance map of the interference pattern overlapping with molecular charge transitions. **b** Derivative of differential conductance ($dI^2_d/dV^2_{sd}$) measured as a function of bias voltage ($V_{sd}$) and gate voltage ($V_g$). **c** Schematic of possible interference conditions. Two possible resonance conditions for Fabry-Pérot interference within the graphene are indicated by arrows within the highly doped region (blue) and on one of the molecule-graphene interfaces (red).

**Graphene-based interference**

A high-resolution conductance map of the $N–2/N–1$ and $N–1/N$ transitions is shown in Fig. 2a. There is periodic structure within the off-resonant conductance, especially in the range ± 100 mV, with a weaker coupling to the gate ($α_{g, FP}$ = 0.054 eV/V) than the molecular states. The effect is more obvious in the second derivative of the current ($dI^2_{sd}/dV^2_{sd}$) map which displays a 'chequerboard' pattern (Fig. 2b).

There are two features of the conductance oscillations that make up the chequerboard, these correspond to two separate periodicities of about 5 and 10 meV (Fig S4-2 displays a two-dimensional fast Fourier transform of the $dI^2_{sd}/dV^2_{sd}$ map). Similar periodicity has been observed in transport measurements through graphene and is attributed to the formation of an electronic Fabry-Pérot (FP) interferometer within the graphene channel.[14,15,31] By analogy with an optical FP cavity formed from a pair of partially reflective mirrors; two potential steps, induced by doping, define an electronic FP cavity. Considering resonances within a one-dimensional FP cavity, our measured periodic energy spacings of 5 and 10 meV correspond to cavity lengths of $L = hv_F/(2E)$ = 1000 nm and 500 nm, using the Fermi velocity of $v_F = 2.4 × 10^6$ m/s for CVD graphene.[31] These length scales are two orders of magnitude larger than those associated with reflections within the **Ni-FP8** nanoribbon itself, indicating the QI pattern is dominated by reflections within the graphene channel, which is further verified by magnetoconductance oscillations (Extended Data Fig. 3).

The cavity is formed within graphene by doping induced by interactions with the substrate. Graphene transferred onto $HfO_2$ is *p*-doped (light purple area,[32] shown schematically in Fig. 2c). The electroburning process that is used to create the nanogap anneals the graphene and increases the hole concentration local to the tunnel junction, generating a highly doped ($p^+$-doped) region (blue area) of ~ 1000 nm in diameter.[31] Therefore reflections over 500 nm occur between a $pp^+$ potential step and a molecule/electrode tunnel barrier, as highlighted by red arrows in Fig. 2c. As the length of the molecule is much shorter than the graphene cavity, reflections from either sides of the molecule-graphene interface are not distinguishable. For the reflections with a 1000 nm cavity (the diameter of the $p^+$-doped region) the interfaces between $p$-doped and $p^+$-doped graphene regions generate the potential steps, the electrons are reflected or transmitted within the $pp^+p$ cavity that contains the molecular junction, as indicated by blue arrows in Fig 2c. As the FP resonances extend over the molecular junction, electron transmission must remain coherent over porphyrin nanoribbon, demonstrating electron transmission through the 8 nm long molecule is phase coherent.

**Phase-coherent molecular transport**

Next, we describe the temperature and charge-state dependence of the device conductance. As shown in Fig. 3a, as the thermal energy increases beyond $hv_F/L$ (5 meV), the FP interference pattern disappears (at 80 K, $k_B T$ = 6.9 meV) likely due to both the thermal energy being larger than the FP spacing, as well as the coherence length within the graphene channel decreasing with temperature.[33] Furthermore, the Kondo resonance, clearly visible as a zero-bias conductance enhancement within the $N$–1 state (**Ni-FP8$^+$**) at 10 K, as shown in Fig. 2a, is also no longer present in the 80 K data. The resonance results from scattering from a many-body state formed from the spin-1/2 radical on the molecule and electrons of opposite spin at the Fermi surface of the electrodes. As thermal energy increases towards the binding energy of this many-body molecule-electrode state the zero-bias conductance will decay. The characteristic temperature-dependence is parameterized by the Kondo temperature ($T_K$), the temperature when the conductance is half its value at 0 K, i.e. G($T$ = $T_K$) = 0.5 × G($T$ = 0). We extract the Kondo temperature of our device at $V_g$ = –0.85 V by fitting the temperature-dependence of the zero-bias conductance to the usual spin-1/2 model: $G_K(T) = G_{T=0}/\left[1 + \left(2^{1/s} - 1\right)\left(T/T_K\right)^2\right]^s$ where $s$ is an empirical parameter.[34] We calculate $T_K$ = 18 ± 1 K and $s$ = 0.30 ± 0.04 from the fit in Fig. 3b. We also obtain $T_K$ at this gate voltage from the FWHM of the resonance that gives a similar value of 17 ± 1 K. (see Fig. 3b, inset). There are several parameters that define Kondo temperature of a nanoscale system; it depends on the addition energy of the $N$–1 transition (0.12 eV for **Ni-FP8**), and exponentially on the molecule-electrode coupling, hence its association with the onset of intermediate coupling.[9] A typical gate-dependent measurement of a Kondo resonance would give a smooth conductance decay (or Kondo valley) as $V_g$ is detuned from the resonances, described by the Haldane relation.[25,34] However, as $T_K$ also depends on the graphene density of states,[35] our gate-dependent measurement of the Kondo peak displays conductance oscillations, as opposed to a smooth valley, within the $N$–1 state due to the interplay between Kondo and FP resonances (visible in the pink gate trace in Fig. 1d, and Supplementary Figure S4-3).

Aside from the Kondo behaviour, the conductance of the device, including at the Coulomb peaks show a weak temperature dependence (Fig. 3c, Fig. 3d). In the regions of $V_g$ between the Coulomb peaks, the conductance is temperature independent, with some slight decrease in the conductance of $N$–1 state and the $N$–1/$N$ peak due to Fermi broadening and consistent with off-resonant phase-coherent transport being the dominant mechanism (more temperature-dependent data are in Supplementary Section 3). This constitutes one of the longest molecular systems (8 nm) with well-defined anchoring over which phase-coherent transport has been measured, resulting from low attenuation factors that are associated with edge-fused porphyrin oligomers.[36]

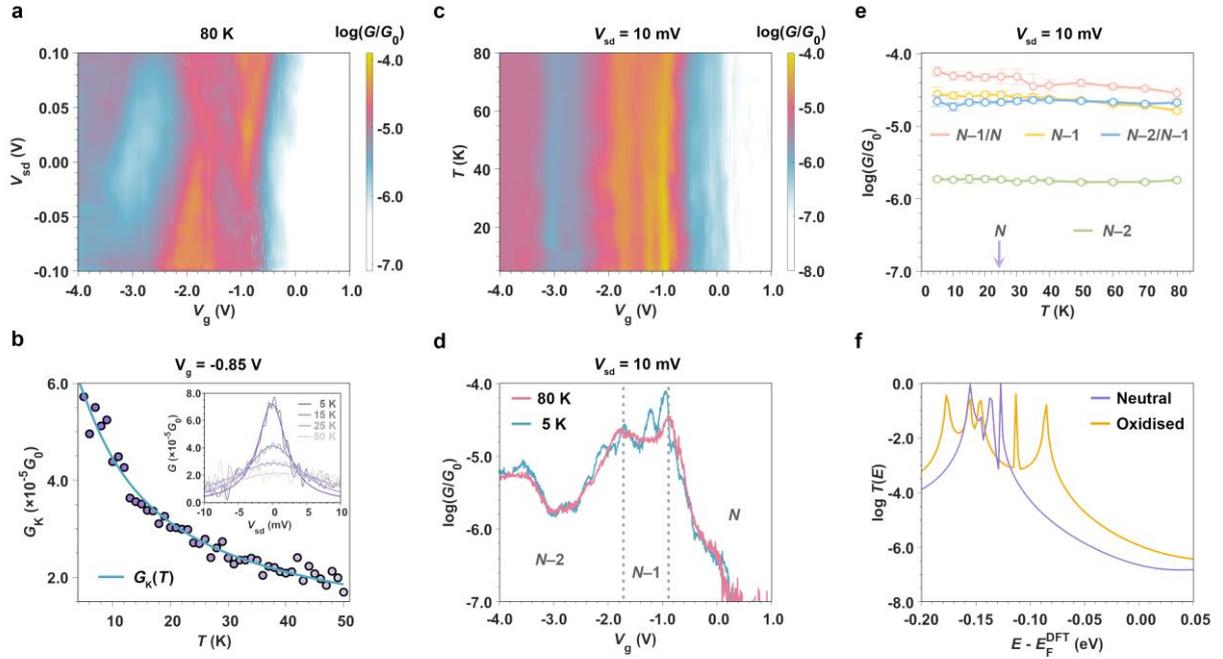

**Fig. 3 | Temperature and oxidation-state dependence of molecular conductance. a** The fluctuations of conductance and Kondo resonances from interference disappear as thermal energy surpasses $hv_F/L$ (5 meV) and the coherence length decreases, shown in the conductance map at 80 K (6.9 meV). **b** Experimental differential conductance of the Kondo peak at $V_g$ = –0.85 V (purple circles) and fit to spin-1/2 Kondo model with $T_K$ = 18 ± 1 K and $s$ = 0.30 ± 0.04 (blue line). Inset: experimental differential conductance as a function of bias voltage (grey curves) and the Lorentzian fits of Kondo peaks (purple curves). **c** Differential conductance map measured as a function of temperature and gate voltage. Indicative of phase coherent transport, the conductance does not shows an obvious temperature dependence. **d** Differential conductance as a function of $V_g$ measured at 5 K (blue curve) and 80 K (pink curve) with a fixed bias voltage $V_{sd}$ = 10 mV. **e** Temperature dependence of off-resonant transport measured in the conductance minima of the charge states. The values are averaged over a 100 mV window around $V_g$ = –2.90 V for $N$–2, $V_g$ = –1.30 V for $N$–1, and $V_g$ = 0.6 V for $N$ charge states. We also display the temperature-dependence of the $N$–2/$N$–1 ($V_g$ = –1.75 V) and $N$–1/$N$ ($V_g$ = –0.95 V) Coulomb peaks. Note there is a shift in the position of the Coulomb peaks from Fig. 2, likely to be a result in a change in trap occupancy in the $HfO_2$. For **c,d** and **e** the conductance was measured $V_{sd}$ = 10 mV to exclude the impact of Kondo resonance. **f** Calculated transmission spectra for the neutral state (purple) and oxidised state (orange) of single-molecule junction.

The data show that off-resonance conductance is strongly dependent on the charge state of **Ni-FP8**, and depends on $V_g$ within each diamond due to the effect of the gate potential on the level alignments (Fig 4d). For the neutral nanoribbon the conductance is below the noise level (< $10^{-7.0}$ $G_0$) of our experimental set-up away from the Coulomb peaks, but it is $10^{-4.7}$ $G_0$ in the mid-gap of the $N$–1 state and $10^{-5.8}$ $G_0$ in $N$–2, indicating that in the oxidised states of the **Ni-FP8** off-resonance coherent transport is more efficient than in the neutral state. Charge-state dependence of conductance has been observed previously in STM-BJ measurements of fused porphyrin oligomers,[37] with a similar conductance enhancement of ~100 reported for an edge-fused trimer upon oxidation, similar to the ratio measured in our device that has a quite different geometry. Furthermore, our results are consistent with the general observation in single-molecule conductance measurements that upon oxidation or reduction, a conductance enhancement is observed in the odd-electron number state.[38-40] As with these previous studies, we can utilize a combination of DFT and quantum transport theory[41-43] to calculate the phase-coherent transmission (see the details in Methods Section) to support our experimental observations of a change in off-resonant conductance after **Ni-FP8** oxidation. To model the effect of a gate potential in the calculations, Cl atoms are placed on top of each Ni atom in the fused octamer to enable transmission to be calculated for an oxidised nanoribbon. The transmission coefficient $T(E)$ (orange curve), when eight Cl atoms are present, is plotted in Fig. 3f and for comparison, the transmission coefficient (purple curve) of the neutral octamer is also shown. In agreement with our measurements, upon oxidation, a clear increase in $T(E)$ is observed over a large energy range close to Fermi energy, primarily due to a shrinking of the HOMO-LUMO gap (Fig. S6-2 in Supplementary Information Section 6). The introduction of Cl atoms does create extra features (spikes) in the transmission that not affect the increasing trend (see Fig. S6-2 in Supplementary Information). The calculated transmission spectrum based on coherent transport theory show excellent agreement with the experimental measurements of **Ni-FP8**, particularly at higher temperature.

## Discussion

Our results show that phase-coherent electron transport persists through a fused porphyrin nanoribbon even over a molecular length of 8 nm. The mechanism is confirmed by conductance oscillations from Fabry-Pérot resonances that extend over a graphene-molecule-graphene cavity and the observation of temperature-independent off-resonant conductance in multiple oxidation states of the molecule. The persistent of phase-coherent transport behaviour across such extended molecular systems, even within the intermediate (rather than strong) molecule-electrode coupling regime, is important for understanding and designing systems for efficient long-range electron transport, and the measurement of molecular conductance in different charge states provides complementary information to measurements of redox switching in single-molecule devices where such control over the molecular oxidation is not possible. The study of coherent electron transport through a molecular nanoribbon embedded in a Fabry-Pérot cavity opens up prospects for all-electrical interferometric measurements between graphene and molecular pathways, where both transmission magnitude and phase through the device can be determined experimentally. This platform could enable the read-out of molecular topological qubits, for which nanoribbons have demonstrated potential,[3] therefore providing an interesting research direction in exploring nanoribbon-graphene hybrid devices for quantum information processing.

## Acknowledgements

This work was supported by the EPSRC (grants EP/N017188/1 and EP/R029229/1) and EU-CoG-MMGNRs. JAM acknowledges funding from the Royal Academy of Engineering and a UKRI Future Leaders Fellowship, Grant No. MR/S032541/1. Part of the substrate fabrication was done at the Center for Nanophase Materials Sciences (CNMS), which is a US Department of Energy, Office of Science User Facility at Oak Ridge National Laboratory.

## Author Contributions

The experiments were conceived by J.O.T. and Z.C. with support from G.A.D.B, C.J.L., J.A.M., L.B., and H.L.A.; Z.C. designed the graphene structure and performed the graphene patterning, and Z.C. and J.O.T. undertook the charge transport measurements. J-R. D. synthesized and characterized the compounds under the supervision of H.L.A.; J.B., X.B. and J.L.S. prepared the device substrates. S.H. and Q.W. performed the DFT calculations of the device under the supervision of C.J.L.; J.O.T. and Z.C. analyzed the data and wrote the paper; all authors discussed the results and edited the manuscript.

## Methods

**Synthesis of Ni-FP8.** To a solution of **Ni-LP8Br** (2.0 mg, 0.15 μmol, 1 equiv.) in dry 1,2-dichloroethane (DCE, 2.5 mL), a suspension of $AuCl_3$ (0.64 mg, 2.1 μmol, 14 equiv.) and AgOTf (2.7 mg, 11 μmol, 70 equiv.) in dry DCE (2.5 mL) was added dropwise and the reaction mixture was stirred at 25 ºC for 15 min. After that, a suspension of $AuCl_3$ (0.13 mg, 0.42 μmol, 2.8 equiv.) and AgOTf (0.54 mg, 2.1 μmol, 14 equiv.) in dry DCE (0.5 mL) was added dropwise to the reaction mixture and the reaction was monitored by UV-vis-NIR spectroscopy with $CH_2Cl_2$ + 1% triethylamine as solvent. After the completion, triethylamine (1.0 mL) was added to the reaction mixture. The resulting mixture was purified by flash column chromatography on silica gel using pentane/$CH_2Cl_2$ (9:1) as eluent to give product **Ni-FP8Br** (1.0 mg, 50% yield).

A mixture of **Ni-FP8Br** (1.0 mg, 0.075 μmol, 1.0 equiv.), Pd(PPh$_3$)$_4$ (1.1 mg, 1.5 μmol, 20 equiv.) and CuI (0.14 mg, 0.75 μmol, 10 equiv.) in dry toluene (0.5 mL) and diisopropylamine (DIPA, 0.5 mL) was degassed by three freeze-pump-thaw cycles. A solution of 1,3,6-tris(dodecyloxy)-8-ethynylpyrene (10 mg, 13 μmol, 170 equiv.) in dry toluene (0.5 mL) and DIPA (0.5 mL) was degassed by three freeze-pump-thaw cycles and transferred to the reaction mixture under argon. After that, the mixture was stirred at 50 ºC under argon for 2 h. Then, a degassed solution of 1,3,6-tris(dodecyloxy)-8-ethynylpyrene (5.0 mg, 6.4 μmol, 85 equiv.) in dry toluene (0.5 mL) and DIPA (0.5 mL) was added to the reaction mixture and the mixture was stirred at 50 ºC for another 20 h. After reaction, the resulting mixture was separated by flash column chromatography on silica gel using pentane/$CH_2Cl_2$ (1:1) as eluent, followed by size-exclusion chromatography (Biorad Bio beads SX-1) with toluene/pyridine (99:1) as eluent to give the crude mixture. The crude mixture was further subjected to recycling GPC with toluene/pyridine (99:1) as eluent to separate the desired product **Ni-FP8** (0.11 mg, 11%). See Extended Data Fig. 1 for the reactions and Supplementary Information Section 2 for characterisation data of intermediate compounds.

**Substrate fabrication.** The substrate used for **Ni-FP8** devices was fabricated using the following procedure. On a degenerately *n*-doped silicon wafer with a layer (300 nm thick) of thermally-grown silicon dioxide ($SiO_2$), a local gate electrode (3 μm wide) was defined by optical lithography with lift-off resist and electron-beam (e-beam) evaporation of titanium (5 nm thick) and platinum (15 nm thick). A layer (10 nm) of hafnium dioxide ($HfO_2$) was then deposited using atomic layer deposition (ALD). Next, source and drain contact electrodes separated by a 7 μm gap (the centre of the gap was aligned to the centre of the gate electrode, which means a 2 μm of horizontal distance between each electrode and gate electrode) were also defined by optical lithography with lift-off resist and electron-beam (e-beam) evaporation of titanium (5 nm thick) and platinum (45 nm thick). The procedure for the fabrication of substrates used for **Zn-P1** and **Zn-FP3** has published previously.[44]

**Graphene nanogaps.** A layer (600 nm) of poly(methyl methacrylate) (PMMA) (with a molecular weight of 495 kDa) was spin coated onto chemical vapour deposition (CVD)-grown graphene (purchased from Grolltex) on copper. The copper was then etched in aqueous ammonium persulfate (($NH_4$)$_2$$S_2$$O_8$) solution (3.6 g in 60 mL water) for 4 hours, after which the PMMA protected graphene was transferred 3 times to Milli-Q water and scooped up using the substrate. Air bubbles were further removed by partly submerging the sample in 2-propanol (IPA). The sample was dried overnight and baked at 180 °C for 1 h. The PMMA was then removed in hot acetone (50 °C) for 3 h.

The Z-shaped graphene tape with bow-tie shaped structure was patterned by e-beam lithography (EBL) with bi-layer lift-off resist (PMMA495 and PMMA950) and thermal evaporation of aluminium (50 nm thick). The Z-shaped graphene pattern was used so the inner graphene leads are coplanar with the bowtie structure (see Fig. 1b, c), reducing tension on the bowtie-shaped graphene, and maximising the stability of the junction. PMMA e-beam resist was used as it is positive resist and it can be transformed into smaller molecules after exposure, which make it much easier to be completely removed than negative photoresist. Aluminium was then deposited onto exposed area as oxygen plasma resist, as aluminium can be completely removed by either acidic or basic aqueous solutions. By this method we reduce contamination from residual photoresist on graphene. The flatter configuration and cleaner surface might provide stronger molecule-electrode coupling by better molecule-graphene interfacing. After lift-off, the graphene on unexposed areas (which are not covered by aluminium) was etched with oxygen plasma. The aluminium was subsequently removed by aqueous sodium hydroxide (NaOH) solution (0.5 M; 1.0 g in 50 mL water). The sample was finally immersed in hot acetone (50 °C) overnight to remove any residual PMMA. The optical image and SEM images can be found in Supplementary Information Section 1. The procedure for the fabrication of substrates used for **Zn-P1** and **Zn-FP3** has published previously.

Graphene nanogaps were prepared by feedback-controlled electroburning of the graphene bow-tie shape until the resistance of the tunnel junction exceeds 1.3 GΩ ($10^{-7}$ $G_0$). The empty nanogaps were characterised by measuring a current map as a function of bias voltage ($V_{sd}$) and gate voltage ($V_g$) at room temperature in order to exclude devices containing residual graphene quantum dots,[16] only clean devices were selected for further measurement.

**Molecule junctions and measurements.** The solution of the porphyrin nanoribbon (1 μM in toluene) was drop-cast on electroburnt graphene electrodes and allowed to dry in air. Only devices that showed clean current maps before molecule deposition were carried on further measurement. Thus, new signals appeared after molecule deposition can be attributed to transport through molecular junctions. Then, the chip contain molecular devices was connected to a chip holder via wire bonding, loaded in Oxford Instruments 4K PuckTester, and cooled down to cryogenic temperature for detailed measurements. The current maps and differential conductance maps of before and after measurements can be found in Supplementary Information Section 3.

**Theoretical Calculations.** Geometrical optimizations were carried out using the DFT code SIESTA,[41] with a local density approximation LDA functional, a double-ζ polarised basis, a cut-off energy of 200 Ry and a 0.04 eV/Å force tolerance. From the Hamiltonian and overlap matrices of the DFT calculation of the junction, Gollum[42] calculates the transmission coefficient $T_{nm}(E)$ between scattering channels n, m in the electrodes, from which the transmission coefficient $T(E) = \sum_{nm} T_{nm}(E)$ is obtained. As discussed in chapter 17 of ref[43], this is equivalent to the expression:

$$T(E) = 4\mathrm{Tr}\left(\Gamma_1 G \Gamma_2 G^\dagger\right)$$

Where $G$ is the (retarded) Green's function of the junction and $\Gamma_i$ is the imaginary part of the self energy of electrode $i$. The electrical conductance is obtained from

$$G = G_0 \int_{-\infty}^{+\infty} T(E) \left(-\frac{\partial}{\partial E} f(E)\right) dE$$

where $E_F$ is the Fermi energy of the device, $f(E) = \dfrac{1}{e^{(E-E_F)/k_B T} + 1}$ is the Fermi distribution function and $G_0 = \dfrac{2e^2}{h}$ is the conductance quantum. At low enough temperatures, this is approximated by $G = G_0 T(E_F)$. In the presence of Cl atoms, spin polarised calculations were carried out to obtain the transmission coefficients $T^\uparrow$, $T^\downarrow$ for the two different spins, from which the total transmission coefficient $T = \dfrac{T^\uparrow + T^\downarrow}{2}$ is obtained.

**Extended Data**

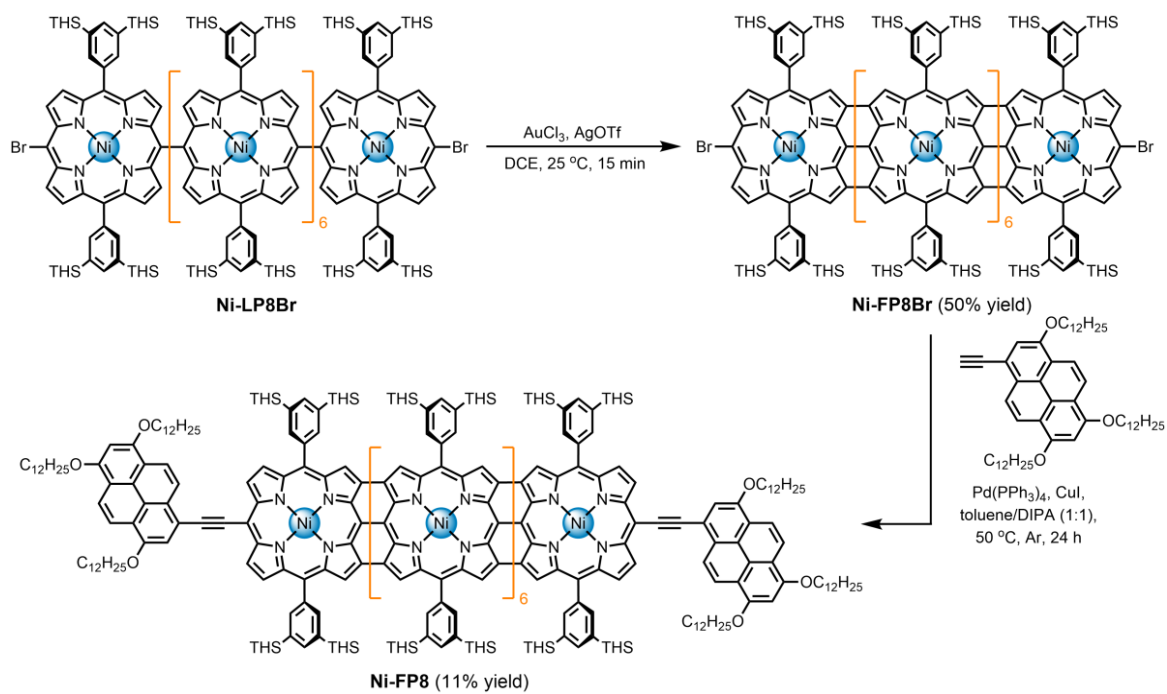

**Extended Data Fig. 1 |** Synthesis of fused porphyrin octamers **Ni-FP8Br** and **Ni-FP8**, THS = trihexylsilyl solubilising group.

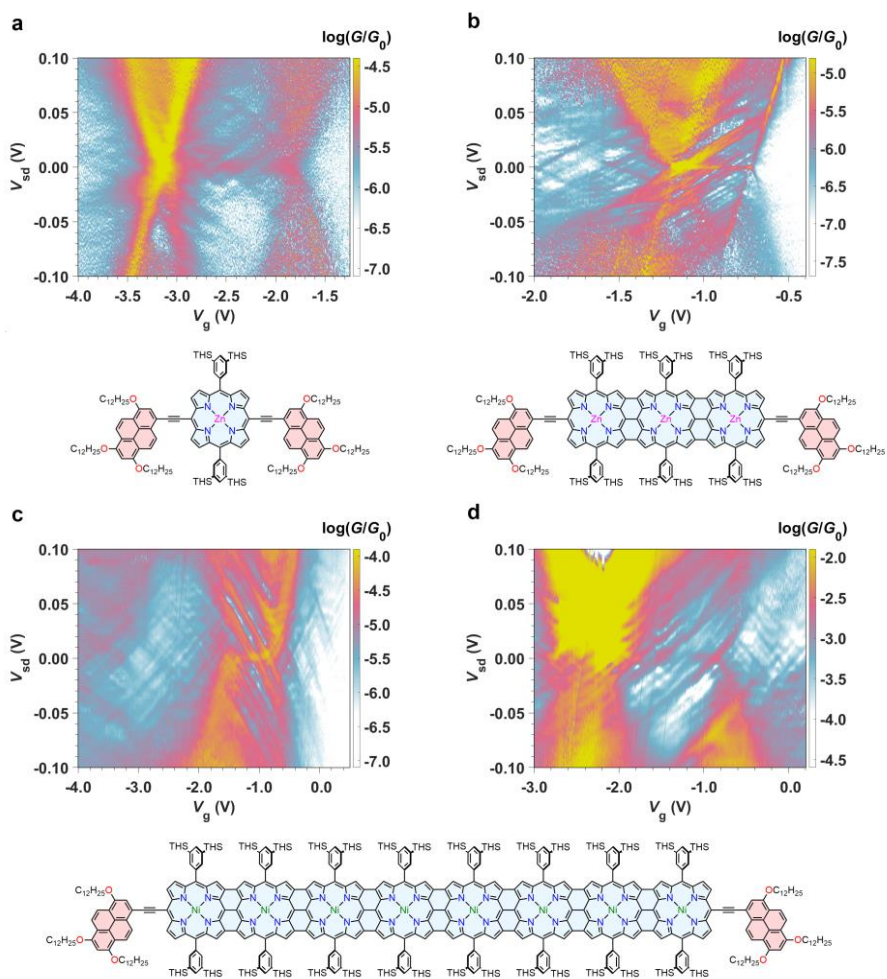

**Extended Data Fig. 2 |** Detailed differential conductance map of the region with interference pattern for devices

with zinc porphyrin monomer (a, **Zn-P1**), with edge-fused zinc porphyrin trimer (b, **Zn-FP3**) and edge-fused nickel porphyrin octamer (c and d, **Ni-FP8**). The **Ni-FP8** device shown in c is discussed in detail in main text.

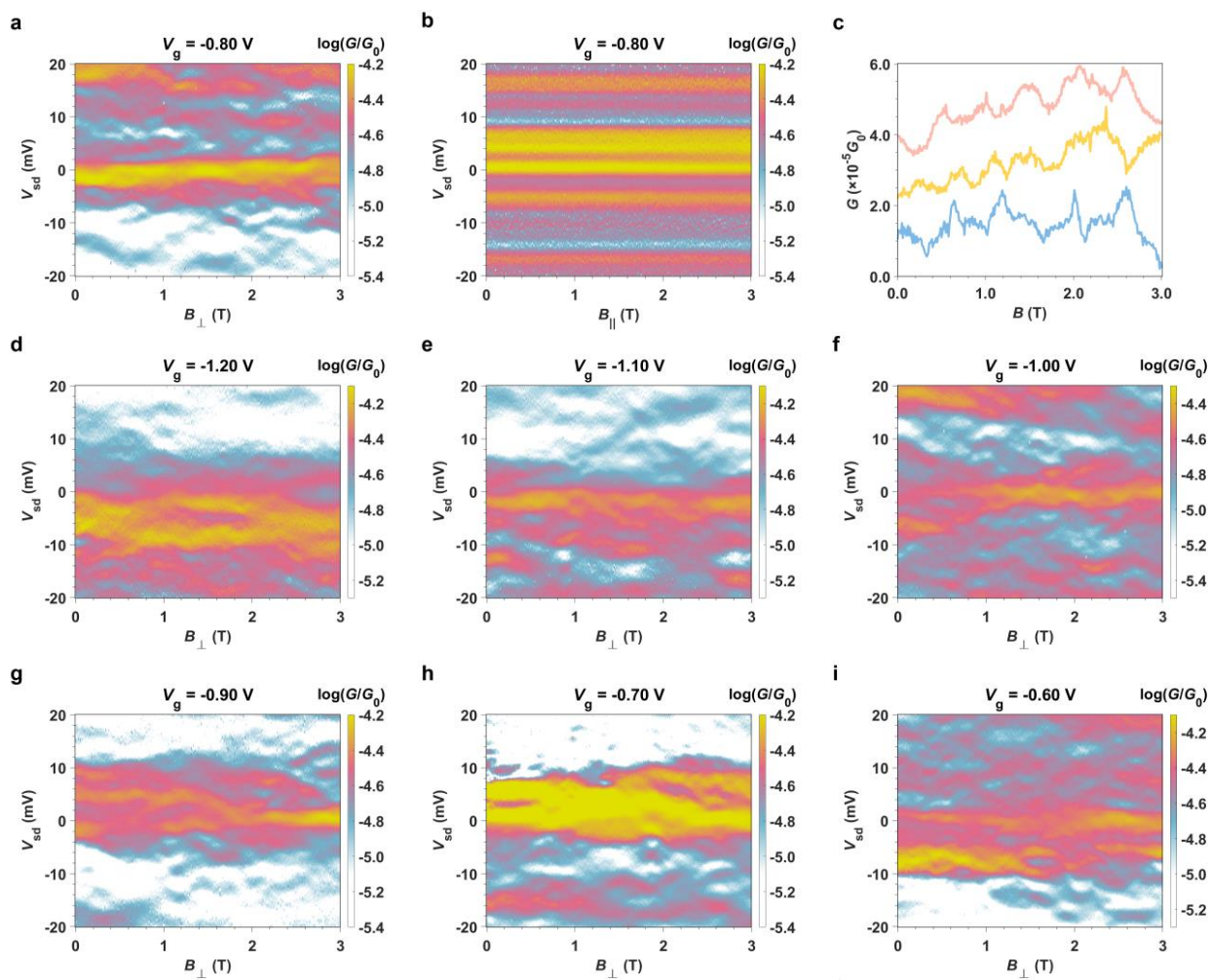

**Extended Data Fig. 3 | a, b** Differential conductance map measured as a function of bias voltage ($V_{sd}$) and magnetic field ($B$) measured at $V_g$ = –0.80 V. The applied magnetic field is perpendicular (⊥, for **a**) and parallel (∥, for **b**) to transport plane (graphene), respectively. The interference pattern changes significantly as increasing perpendicular magnetic field, but unchanged in parallel magnetic field. **c** Weak magnetoconductance oscillations at $V_{sd}$ = –5 mV (blue), 1 mV (yellow) and 9 mV (red) with a period of ~ 0.5 T (adjacent traces are offset for clarity), the origins of which are discussed in the Supplementary Information Section 5. **d-i** Differential conductance map measured at different $V_g$ under perpendicular (⊥) magnetic field. The oscillations on interference patterns are significant. Under parallel magnetic no obvious change was observed at all $V_g$, which are similar with that shown in **b**. Further details can be found in Supplementary Information Section 5.